\documentclass{llncs}
\usepackage{llncsdoc}

\usepackage{amsmath}
\usepackage{color}
\newcommand{\tr}[1]{\textcolor{red}{#1}}

\def\R{{\bbbr}}
\begin{document}
\title{A  Simple Linear Ranking Algorithm Using Query Dependent Intercept Variables}
\author{\vspace{-1ex}Nir Ailon}
\institute{\vspace{-1ex}Google Research, 76 Ninth Ave, 4th Floor, New York NY 10011}
\maketitle
\begin{abstract}
\vspace{-5ex}
The LETOR website   
contains three information retrieval datasets used as a benchmark for testing machine learning ideas
for ranking.  Algorithms participating in the challenge
are required to assign score values to  search results for a collection of queries, and are measured using standard IR ranking measures
(NDCG, precision, MAP) that depend only the relative score-induced order of the results.
Similarly to many of the ideas proposed in the participating algorithms, we train a linear
classifier.  In contrast with other participating algorithms, we define an additional free variable (intercept, or benchmark)
for each query.  This allows expressing the fact that results
for different queries are incomparable for the purpose of determining relevance.  The cost of this idea is the addition
of relatively few nuisance parameters.  Our approach is simple,  and we used 
a standard logistic regression library to test it.  The results beat the reported participating algorithms.
Hence, it seems promising to combine our approach with other more complex ideas.
\end{abstract}

\vspace{-6ex}
\section{Introduction}
\vspace{-2ex}
The LETOR benchmark dataset \cite{LQXXL}   {\tt http://research.microsoft.com/users/LETOR/}  (version 2.0) contains three information retrieval datasets used as a benchmark for testing machine learning ideas
for ranking. Algorithms participating in the challenge
are required to assign score values to  search results for a collection of queries, and are measured using standard IR ranking measures (NDCG@$n$, precision@$n$ and MAP - see \cite{LQXXL} for details), designed in such a way that only the relative order of the results matters.
The input to the learning problem is a list of query-result records, where each record is a vector of standard IR features
together with a relevance label and a query id.  The label is either binary (irrelevant or relevant) or trinary (irrelevant, relevant or very relevant).
%and was obtained by combining several judges' responses. 

All reported algorithms used for this task on LETOR website \cite{CQLTL07,FSS03,HGO99,QZWLLL07,TLQCM07,XL07} 
rely on the fact that records corresponding to the same query id are
in some sense comparable to each other, and cross query records are incomparable.  The rationale is that the
IR measures are computed as a sum over the queries, where for each query  a nonlinear function is computed.
For example, RankSVM \cite{HGO99} and RankBoost \cite{FSS03} use pairs of results for the same query to
penalize a cost function, but never cross-query pairs of results.
%The nonlinear part depends only the permutation induced on the corresponding query results by the algorithm scores.

The following approach seems at first too naive compared to others: Since the training  information is given as relevance labels,
 why not simply train a linear classifier to predict the relevance labels, and use prediction confidence  as score?
  Unfortunately this approach fares poorly.  The hypothesized reason is that judges' relevance response may
depend on the query.  To check this hypothesis, we define an additional free variable (\emph{intercept} or \emph{benchmark})
for each query.  This allows expressing the fact that results
for different queries are incomparable for the purpose of determining relevance.  The cost of this idea is the addition
of relatively few nuisance parameters.  Our approach is extremely simple, and we used
a standard logistic regression library to test it on the data.  
This work is not the first to suggest query dependent ranking, but it is arguably the simplest, most immediate way to address
this dependence using linear classification before other complicated ideas should be tested.
Based on our judgment, other reported algorithms used for the challenge are more complicated, and our
solution is overall better on the given data.

%The purpose of this work is to show that since the input to the problem is in the form of labels on individual
%query-result records, it makes sense to simply learn how to predict these labels, and use the prediction confidence
%level as a score for sorting results.
% Unfortunately if such an approach is
%performed naively then the results thus obtained are poor.  The problem is that this approach does not take into
%account that the criteria used for labeling results depends on the context, viz, the query 
 %The techniques used to circumvent this problem in previous work (as reported on the website) do so indirectly.
% For example, RankSVM trains a linear classifier using pairs of results for single queries as training samples.
 %In this work we also train a linear classifier, and we address 
 %ake a more direct approach
% Here we take the following
% However, if a different intercept variable is assigned
%to each query, then the results are competitive with the best known results obtained for the problem.  Assigning a
%different intercept variable to different queries is a simple way to encode the fact that the relevance criterion may vary across
%different queries.

\vspace{-3ex}
\section{Theory and Experiments}
\vspace{-2ex}
\def\O{{\mathcal O}}
\def\T{{\mathcal T}}
\def\tree{\operatorname{tree}}
Let $Q_i$, $i=1,\dots,n$ be a sequence of queries, and for each $i$ let $R_{i1},\dots, R_{im_i}$ denote a corresponding
set of retrieved results.  For each $i\in [n]$ and $j\in [m_i]$ let $\Phi_{ij}  = (\Phi_{ij}(1),\dots \Phi_{ij}(k))\in \R^k$ denote a real valued feature vector.  
Here,  the coordinates of $\Phi_{ij}$ are standard IR features.  Some of these features depend on the result only,
and some on the query-result pair, as explained in \cite{LQXXL}.  
Also assume that for each $i,j$ there is a judge's response label $L_{ij} \in \O$,
where $\O$ is a finite set of ordinals.  In the TREC datasets (TD2003 and TD2004), $\O=\{0,1\}$. In the OHSUMED dataset $\O=\{0,1,2\}$.    Higher numbers
represent higher relevance.

{\bf The Model.}
We assume the following generalized linear model for $L_{ij}$ given $\Phi_{ij}$ using the logit link.  Other models are possible,
but we chose this one for simplicity.    Assume first that the set of ordinals is binary: $\O=\{0,1\}$.
  There is a hidden global weight vector $w\in \R^k$.
%as well as an ordered binary tree structure $\T$ with the ordinals $\O$ as leaves.  In the cases we consider there aren't many possibilities
%for $\T$.  In the TREC case, $\T = \tree(0,1)$, and in OHSUMED we either have $\T=\tree(0,\tree(1,2))$ or $\T=\tree(\tree(0,1),2)$.
%For semantic reasons we choose the former case, because $1$ and $2$ both express positive relevance, whereas $0$ expresses
%irrelevance. 
% The exact representation of $\O$ as a tree is not the main point of this paper and is chosen so that classification
%in the trinary case is done as a sequence of  two binary classifications.
Aside from $w$, there is a query dependent parameter $\Theta_i\in \R$ corresponding to each query $Q_i$.  We call this
parameter a \emph{benchmark} or an \emph{intercept}.   The intuition behind defining this parameter is to allow for a different relevance criterion to different queries.
The probability distribution $\Pr_{w,\Theta_i}(L_{ij}| Q_i,R_{ij})$ of response to result $j$ for query $i$ is given by%Let $u$ denote a leaf of $\T$ and assume
%$v_1,\dots,v_{l_u}$ is the root-to-leaf path to $u$ in $\T$.  Then
%$$ \Pr_{w,\T,\Theta}(L_{ij}=u|Q_i,R_{ij}) = \prod_{r=1}^{l_u} \frac 1 {1+e^{(w\cdot \Phi_{ij} - \Theta_i(v_r))[u \succ_\T v_r]}}\ , $$

\vspace{-3ex}
$$ \Pr_{w,\Theta_i}(L_{ij}=1|Q_i,R_{ij}) = \frac 1 {1+e^{\Theta_i - w\cdot \Phi_{ij} }} \ \ \  \Pr_{w,\Theta_i}(L_{ij}=0|Q_i,R_{ij}) = \frac 1 {1+e^{w\cdot \Phi_{ij} - \Theta_i }} $$
\noindent
\vspace{-3ex}

In words, the probability of result $j$ for query $i$ deemed relevant is $\Theta_i - w\cdot \Phi_{ij}$ passed through the logit link, where
$w\cdot \Phi_{ij}$ is vector dot product.
This process should be thought of as a statistical comparison between  the value of a search result $R_{ij}$ (obtained as a
linear function of its feature vector $\Phi_{ij}$) to a benchmark $\Theta_i$.  In our setting, both the linear coefficients
$w$ and the benchmark $\Theta_1,\dots, \Theta_n$ are variables
which can  be efficiently learnt in the maximum likelihood (supervised) setting.  Note that the total 
number of variables is $n$ (number of queries) plus $k$ (number of features).  

{\bf Observation: } For  any weight vector $w$, benchmark variable $\Theta_i$ corresponding to query $Q_i$ and two result  incides $j,k$,
\vspace{-1ex}
$$ \Pr_{w,\Theta_i}(L_{ij}=1 | Q_i,R_{ij}) >  \Pr_{w,\Theta_i}(L_{ik}=1 | Q_i,R_{ik}) \iff w\cdot \Phi_{ij} > w\cdot \Phi_{ik}\ . $$
This last observation means that for the purpose of ranking candidate results for a specific query $Q_i$ in decreasing order
of relevance likelihood, the benchmark parameter $\Theta_i$  is not needed.  Indeed, in our experiments below the
benchmark variables will be used only in conjunction with the training data.  In testing, this variable will neither be known nor necessary.

\noindent
{\bf The Trinay Case.}
As stated above, the labels for the OHSUMED case are trinary: $\O=\{0,1,2\}$.  We chose the following model to extend the binary case.  Instead of one benchmark parameter for each query  $Q_i$ there are two such parameters, $\Theta_i^H, \Theta_i^L$ (\emph{H}igh/ \emph{L}ow) with $\Theta_i^H > \Theta_i^L$.    Giver a candidate result $R_{ij}$ to query $Q_i$ and the parameters, the probability distribution on 
the three possible ordinals is:
\vspace{-4ex}
%\begin{equation}
$$\Pr_{w,\Theta_i^H,\Theta_i^L}(L_{ij} = X | Q_i, R_{ij}) = 
\begin{cases} 
\frac {1}  {\left (1+e^{w\cdot \Phi_{ij} - \Theta^H_i}\right )   \left (1+e^{w\cdot \Phi_{ij} - \Theta^L_i}\right )} & X=0 \\
\frac 1  { \left(1+e^{w\cdot \Phi_{ij} - \Theta^H_i}\right )   \left (1+e^{\Theta^L_i - w\cdot \Phi_{ij}}\right)} & X=1 \\
\frac 1  { \left(1+e^{\Theta^H_i - w\cdot \Phi_{ij} }\right ) } & X=2 \\
\end{cases}
$$
%\end{equation}

\vspace{-2ex}
In words, the result $R_{ij}$ is statistically compared against benchmark $\Theta_i^H$.  If it is deemed higher than the benchmark,
the label $2$ ("very relevant") is outputted as response.  Otherwise, the result is statistically compared against benchmark $\Theta_i^L$,
and the resulting comparison is either $0$ (irrelevant) or $1$ (relevant).\footnote{A natural alternative to this model is the following:  Statistically compare against $\Theta_i^L$ to decide of the result is irrelevant.
If it is not irrelevant, compare against $\Theta_i^H$ to decide between relevant and very relevant.  In practice, the model proposed
above gave better results.}
The model is inspired by Ailon and Mohri's QuickSort algorithm, proposed as a learning method in their recent paper \cite{AM07}:
Pivot elements (or, benchmarks) are used to  iteratively refine the ranking of data. 
%Note that this work is not directly related to their result because (i) In their work, QuickSort is not used to define a ranking
%model, but rather as a technique for reducing binary classification to ranking.  (ii) In their work, the benchmarks are repeatedly chosen at random
%from within the input elements until a full ranking is obtained, whereas here the benchmark is an external unknown parameter, and the
%pivoting is done only twice:  Once to determine the top level comparisons, and a second time to further refine the comparisons
%of the losers in the top level.
%are $k=\lceil \log_2|\O|\rceil$ hidden parameter $\Theta_{i1},\dots,\Theta_{ik}$ which we call a benchmark variables.  

%We use logistic regression in this work, although other classification techniques can be used as well.
%In the first attempt, we assume the following generative model for determining binary relevance labels.

%The underlying generative model for binary relevance labels is assumed:  The query-result is encoded as a $k$ dimensional
%real valued vector.   There is a global weight  vector $v\in \R^k$ and a 
% The labeler computes the dot product of $v\in \R^k$ with a fixed weight vector $w\in \R^k$.  
  \vspace{-2ex}
 \noindent
 \\
 {\bf Experiments.}
 We used an out of the box implementation of logistic regression in R to test the above ideas.  Each one of the three datasets
 includes $5$ folds of data, each fold consisting of training, validation (not used) and testing data.   
From each training dataset, the variables $w$ and $\Theta_i$ (or $w, \Theta_i^H, \Theta_i^L$ in the OHSUMED case) were recovered in the maximum likelihood sense (using logistic regression).  Note that the constraint $\Theta_i^H > \Theta_i^L$ was not enforced,
but was obtained as a byproduct.   The weight vector $w$ was then used to score the test data.  The scores were
 passed through an evaluation tool provided by the LETOR website.  % consisting of three standard IR ranking score types: NDCG, precision
 %and MAP.    The results, summarized in the following section, are comparable to the currently winning techniques. 

 \vspace{-2ex}
\noindent
\\
 {\bf Results.}
The results for OHSUMED are summarized in Tables~\ref{OHSUMEDNDCG},~\ref{OHSUMEDprecision}, and ~\ref{MAP}.
The results for TD2003 are summarized in Tables~\ref{TD2003NDCG},~\ref{TD2003precision}, and~\ref{MAP}.
The results for TD2004 are summarized in Tables~\ref{TD2004NDCG},~\ref{TD2004precision}, and~\ref{MAP}.
The significance of each score separately is
quite small (as can be seen by the standard deviations), but it is clear that overall our method outperforms the others.  For
convenience, the winning average score (over 5 folds) is marked in red for each table column.

 \begin{table}[h!]
 \begin{tabular}{|c|c|c|c|c|c|}
 \hline
  & @2 & @4 & @6 & @8 &  @10 \\ 
  \hline
 This &  \textcolor{red}{$0.491 \pm 0.086$} & \textcolor{red}{$0.480 \pm 0.058$} & \textcolor{red}{$0.458 \pm 0.055$} & \textcolor{red}{$0.448 \pm 0.054$}& $0.447 \pm 0.047$  \\
  RankBoost & $0.483 \pm 0.079$ & $0.461 \pm 0.063$ & $0.442 \pm 0.058$ & $0.436 \pm 0.044$ & $0.436 \pm 0.042$ \\
  RankSVM & $0.476 \pm 0.091$ & $0.459 \pm 0.059$ & $0.455 \pm 0.054$ & $0.445 \pm 0.057$ & $0.441 \pm 0.055$ \\
   FRank & $0.510 \pm 0.074$ & $0.478 \pm 0.060$ & $0.457 \pm 0.062$ & $0.445 \pm 0.054$ & $0.442 \pm 0.055$  \\
 ListNet & $0.497 \pm 0.062$ & $0.468 \pm 0.065$ & $0.451 \pm 0.056$ & $0.451 \pm 0.050$ & \textcolor{red}{$0.449 \pm 0.040$}   \\
 AdaRank.MAP &  $0.496 \pm 0.100$ & $0.471 \pm 0.075$ & $0.448 \pm 0.070$ & $0.443 \pm 0.058$ & $0.438 \pm 0.057$ \\
 AdaRank.NDCG & $0.474 \pm 0.091$ & $0.456 \pm 0.057$ & $0.442 \pm 0.055$ & $0.441 \pm 0.048$ & $0.437 \pm 0.046$ \\
 \hline
 \end{tabular}
 \caption{OHSUMED:  Mean $\pm$ Stdev for NDCG over 5 folds} \label{OHSUMEDNDCG}
 \vspace{-5ex}
 \end{table}

 \begin{table}[h!]
 \begin{tabular}{|c|c|c|c|c|c|}
 \hline
 & @2 & @4 & @6 & @8 &  @10 \\ 
  \hline
 This &  $0.610 \pm 0.092$ & \textcolor{red}{$0.598 \pm 0.082$} & \textcolor{red}{$0.560 \pm 0.090$} & \textcolor{red}{$0.526 \pm 0.092$} &\textcolor{red}{$0.511 \pm 0.081$} \\
  RankBoost & $0.595 \pm 0.090$ & $0.562 \pm 0.081$ & $0.525 \pm 0.093$ & $0.505 \pm 0.072$ & $0.495 \pm 0.081$ \\
  RankSVM &  $0.619 \pm 0.096$ & $0.579 \pm 0.072$ & $0.558 \pm 0.077$ & $0.525 \pm 0.088$ & $0.507 \pm 0.096$ \\
   FRank & $0.619 \pm 0.051$ & $0.581 \pm 0.079$ & $0.534 \pm 0.098$ & $0.501 \pm 0.091$ & $0.485 \pm 0.097$ \\
 ListNet & \textcolor{red}{$0.629 \pm 0.080$} & $0.577 \pm 0.097$ & $0.544 \pm 0.098$ & $0.520 \pm 0.098$ & $0.510 \pm 0.085$ \\
 AdaRank.MAP & $0.605 \pm 0.102$ & $0.567 \pm 0.087$ & $0.528 \pm 0.102$ & $0.502 \pm 0.087$ & $0.491 \pm 0.091$ \\
 AdaRank.NDCG & $0.605 \pm 0.099$ & $0.562 \pm 0.063$ & $0.529 \pm 0.073$ & $0.506 \pm 0.073$ & $0.491 \pm 0.082$ \\
 \hline
 \end{tabular}
 \caption{OHSUMED: Mean $\pm$ Stdev for precision over 5 folds}\label{OHSUMEDprecision}
  \vspace{-5ex}
 \end{table}

% \begin{table}[h!]
% \begin{center}
% \begin{tabular}{|c|c|}
% \hline
%  & MAP \\
%  \hline
%  This & $0.445 \pm 0.065$ \\
%  RankBoost & $0.440 \pm 0.062$ \\
%    RankSVM & $0.447 \pm 0.067$ \\
%      FRank & $0.446 \pm 0.062$ \\
%  ListNet & $0.450 \pm 0.063$ \\
%  AdaRank.MAP & $0.442 \pm 0.061$ \\
%  AdaRank.NDCG & $0.442 \pm 0.058$ \\
%    \hline
% \end{tabular}
% \end{center}
% \caption{OHSUMED:  Mean $\pm$ Stdev for MAP over 5 folds}\label{OHSUMEDMAP}
% \end{table}

 \begin{table}[h!]
 \begin{tabular}{|c|c|c|c|c|c|}
 \hline
  & @2 & @4 & @6 & @8 &  @10 \\ 
  \hline
 This & \textcolor{red}{$0.430 \pm 0.179$} & \textcolor{red}{$0.398 \pm 0.146$} & $0.375 \pm 0.125$ & $0.369 \pm 0.113$ & $0.360 \pm 0.105$ \\
  RankBoost & $0.280 \pm 0.097$ & $0.272 \pm 0.086$ & $0.280 \pm 0.071$ & $0.282 \pm 0.074$ & $0.285 \pm 0.064$ \\ 
  RankSVM &$0.370 \pm 0.130$ & $0.363 \pm 0.132$ & $0.341 \pm 0.118$ & $0.345 \pm 0.117$ & $0.341 \pm 0.115$ \\
   FRank & $0.390 \pm 0.143$ & $0.342 \pm 0.107$ & $0.330 \pm 0.087$ & $0.332 \pm 0.079$ & $0.336 \pm 0.074$ \\
 ListNet & \textcolor{red}{$0.430 \pm 0.160$} &$0.386 \pm 0.125$ & \textcolor{red}{$0.386 \pm 0.106$} & \textcolor{red}{$0.373 \pm 0.104$} & \textcolor{red}{$0.374 \pm 0.094$} \\
 AdaRank.MAP & $0.320 \pm 0.104$ & $0.268 \pm 0.120$ & $0.229 \pm 0.104$ & $0.206 \pm 0.093$ & $0.194 \pm 0.086$ \\
 AdaRank.NDCG & $0.410 \pm 0.207$ & $0.347 \pm 0.195$ & $0.309 \pm 0.181$ & $0.286 \pm 0.171$ & $0.270 \pm 0.161$ \\
 \hline
 \end{tabular}
 \caption{TD2003:  Mean $\pm$ Stdev for NDCG over 5 folds}\label{TD2003NDCG}
  \vspace{-5ex}
 \end{table}

 \begin{table}[h!]
 \begin{tabular}{|c|c|c|c|c|c|}
 \hline
 & @2 & @4 & @6 & @8 &  @10 \\ 
  \hline
 This & \textcolor{red}{$0.420 \pm 0.192$} & \textcolor{red}{$0.340 \pm 0.161$} & \textcolor{red}{$0.283 \pm 0.131$} & \textcolor{red}{$0.253 \pm 0.115$} & \textcolor{red}{$0.222 \pm 0.106$} \\
  RankBoost &$0.270 \pm 0.104$ & $0.230 \pm 0.112$ & $0.210 \pm 0.080$ & $0.193 \pm 0.071$ & $0.178 \pm 0.053$ \\
  RankSVM &  $0.350 \pm 0.132$ & $0.300 \pm 0.137$ & $0.243 \pm 0.100$ & $0.233 \pm 0.091$ & $0.206 \pm 0.082$ \\
   FRank & $0.370 \pm 0.148$ & $0.260 \pm 0.082$ & $0.223 \pm 0.043$ & $0.210 \pm 0.045$ & $0.186 \pm 0.049$  \\
 ListNet & \textcolor{red}{$0.420 \pm 0.164$} & $0.310 \pm 0.129$ & \textcolor{red}{$0.283 \pm 0.090$} & $0.240 \pm 0.075$ & \textcolor{red}{$0.222 \pm 0.061$} \\
 AdaRank.MAP & $0.310 \pm 0.096$ & $0.230 \pm 0.105$ & $0.163 \pm 0.081$ & $0.125 \pm 0.064$ & $0.102 \pm 0.050$ \\
 AdaRank.NDCG &$0.400 \pm 0.203$ & $0.305 \pm 0.183$ & $0.237 \pm 0.161$ & $0.190 \pm 0.140$ & $0.156 \pm 0.120$ \\
 \hline
 \end{tabular}
 \caption{TD2003: Mean $\pm$ Stdev for precision over 5 folds}\label{TD2003precision}
  \vspace{-5ex}
 \end{table}

 \begin{table}[h!]
 \begin{tabular}{|c|c|c|c|c|c|}
 \hline
  & @2 & @4 & @6 & @8 &  @10 \\ 
  \hline
 This &\textcolor{red}{$0.473 \pm 0.132$} & \textcolor{red}{$0.454 \pm 0.075$} & \textcolor{red}{$0.450 \pm 0.059$} & $0.459 \pm 0.050$ & \textcolor{red}{$0.472 \pm 0.043$} \\
  RankBoost &\textcolor{red}{$0.473 \pm 0.055$} & $0.439 \pm 0.057$ & $0.448 \pm 0.052$ & \textcolor{red}{$0.461 \pm 0.036$} & \textcolor{red}{$0.472 \pm 0.034$} \\
  RankSVM & $0.433 \pm 0.094$ & $0.406 \pm 0.086$ & $0.397 \pm 0.082$ & $0.410 \pm 0.074$ & $0.420 \pm 0.067$ \\
   FRank & $0.467 \pm 0.113$ & $0.435 \pm 0.088$ & $0.445 \pm 0.078$ & $0.455 \pm 0.055$ & $0.471 \pm 0.057$ \\
 ListNet & $0.427 \pm 0.080$ & $0.422 \pm 0.049$ & $0.418 \pm 0.057$ & $0.449 \pm 0.041$ & $0.458 \pm 0.036$ \\
 AdaRank.MAP & $0.393 \pm 0.060$ & $0.387 \pm 0.086$ & $0.399 \pm 0.085$ & $0.400 \pm 0.086$ & $0.406 \pm 0.083$ \\
 AdaRank.NDCG &$0.360 \pm 0.161$ & $0.377 \pm 0.123$ & $0.378 \pm 0.117$ & $0.380 \pm 0.102$ & $0.388 \pm 0.093$ \\
 \hline
 \end{tabular}
 \caption{TD2004:  Mean $\pm$ Stdev for NDCG over 5 folds}\label{TD2004NDCG}
  \vspace{-8ex}
 \end{table}

 \begin{table}[h!]
 \begin{tabular}{|c|c|c|c|c|c|}
 \hline
 & @2 & @4 & @6 & @8 &  @10 \\ 
  \hline
 This & \tr{$0.447 \pm 0.146$} & \tr{$0.370 \pm 0.095$} & \tr{$0.316 \pm 0.076$} & \tr{$0.288 \pm 0.076$} & \tr{$0.264 \pm 0.062$} \\
  RankBoost & \tr{$0.447 \pm 0.056$}& $0.347 \pm 0.083$ & $0.304 \pm 0.079$ & $0.277 \pm 0.070$ & $0.253 \pm 0.067$ \\
  RankSVM & $0.407 \pm 0.098$ & $0.327 \pm 0.089$ & $0.273 \pm 0.083$ & $0.247 \pm 0.082$ & $0.225 \pm 0.072$ \\
   FRank & $0.433 \pm 0.115$ & $0.340 \pm 0.098$ & $0.311 \pm 0.082$ & $0.273 \pm 0.071$ & $0.256 \pm 0.071$ \\
 ListNet & $0.407 \pm 0.086$ & {$0.357 \pm 0.087$} & $0.307 \pm 0.084$ & $0.287 \pm 0.069$ & $0.257 \pm 0.059$ \\
 AdaRank.MAP & $0.353 \pm 0.045$ & $0.300 \pm 0.086$ & $0.282 \pm 0.068$ & $0.242 \pm 0.063$ & $0.216 \pm 0.064$ \\
 AdaRank.NDCG & $0.320 \pm 0.139$ & $0.300 \pm 0.082$ & $0.262 \pm 0.092$ & $0.232 \pm 0.086$ & $0.207 \pm 0.082$ \\
 \hline
 \end{tabular}
 \caption{TD2004: Mean $\pm$ Stdev for precision over 5 folds}\label{TD2004precision}
  \vspace{-5ex}
 \end{table}

 \begin{table}[h!]
\begin{center}
 \begin{tabular}{|c|c|c|c|} % c|}
 \hline
  & OHSUMED &TD2003 & TD2004 \\ %& Total \\
  \hline
  This & $0.445 \pm 0.065$ & $0.248 \pm 0.075$ &$0.379 \pm 0.051$  \\ % &  $1.072 \pm 0.112$ \\
  RankBoost &  $0.440 \pm 0.062$ &  $0.212 \pm 0.047$ & \tr{$0.384 \pm 0.043$}  \\ % &  $1.036 \pm 0.089$ \\
    RankSVM & {$0.447 \pm 0.067$} & {$0.256 \pm 0.083$} & $0.350 \pm 0.072$ \\ %& $1.053 \pm 0.129$\\
      FRank &$0.446 \pm 0.062$ &$0.245 \pm 0.065$ &  $0.381 \pm 0.069$ \\ %& $1.072 \pm  0.113$ \\
  ListNet & \tr{$0.450 \pm 0.063$}  & \tr{$0.273 \pm 0.068$}  & $0.372 \pm 0.046$ \\ %& $1.095 \pm  0.103$ \\
  AdaRank.MAP & $0.442 \pm 0.061$ &$0.137 \pm 0.063$ &$0.331 \pm 0.089$ \\ % &  $0.910 \pm 0.125$ \\
  AdaRank.NDCG & $0.442 \pm 0.058$ &$0.185 \pm 0.105$ &  $0.299 \pm 0.088$  \\ %& $ 0.926 \pm 0.149$\\
    \hline
 \end{tabular}
 \end{center}
 \vspace{-2ex}
 \caption{Mean $\pm$ Stdev for MAP over 5 folds}\label{MAP}
  \vspace{-6ex}
 \end{table}
 \noindent

\noindent
{\bf Conclusions and further ideas}
$\bullet$ In this work we showed that a simple out-of-the-box generalized linear model using logistic regression performs as least as well
 the state of the art in learning ranking algorithms if a separate intercept variable (benchmark) is defined for each query
 $\bullet$ In a more
 eleborate IR system, a separate intercept variable could be attached to each pair of  \emph{query} $\times$ \emph{judge}
 (indeed, in LETOR the separate judges' responses were aggregated somehow, but in general it is likely that different judges would
 have different benchmarks as well) $\bullet$  The simplicity of our approach is also its main limitation.  However, 
it  can easily be implemented in conjunction with other
 ranking ideas.  For  example, recent work by Geng et al. \cite{GLQLS} (not evaluated on LETOR) proposes query dependent ranking, where
 the category  of a query is determined using a k-Nearest Neighbor method.  It is immediate to apply the ideas here within
each category.

 \bibliographystyle{plain}
 \vspace{-3ex}
\bibliography{LETOR}   

 \end{document}